\documentclass{article}
\usepackage{authblk}
\usepackage{geometry}
\usepackage{hyperref}
\usepackage{tikz}
\usepackage{graphicx}
\usetikzlibrary{shapes.geometric, arrows.meta, positioning, fit, backgrounds}
\usepackage{booktabs} 
\usepackage{caption}  
\usepackage{array}    
\usepackage[utf8]{inputenc}
\title{\textit{Million Tutoring Moves (MTM)}: An Open Multimodal Dataset for the Science of Tutoring}

\author[1,3]{René Kizilcec}
\author[1,3]{Kirk Vanacore}
\author[1,3]{Zhuqian Zhou}
\author[1,2]{Doug Pietrzak}
\author[1,2]{Jorge Dias}
\author[1,3]{Haocheng Zhang}
\author[1,3]{Bakhtawar Ahtisham}
\author[1,3]{Joshua Marland}
\author[1,3]{Rachel Slama}
\author[1,4]{Justin Reich}
\author[1,5]{Kenneth Koedinger}

\affil[1]{National Tutoring Observatory}
\affil[2]{FreshCognate}
\affil[3]{Cornell University}
\affil[4]{Massachusetts Institute of Technology}
\affil[5]{Carnegie Mellon University}
\date{2026}

\begin{document}

\maketitle

\begin{abstract}
We introduce the Million Tutoring Moves (MTM) project, an open dataset initiative aimed at advancing the science of tutoring through large-scale, reusable, and multimodal interaction data. MTM is developed within the National Tutoring Observatory (NTO), a research infrastructure designed to study authentic tutoring interactions and translate them into actionable insights for research, practice, and AI-powered educational technology development. In this paper, we present the vision behind MTM and describe MTM v1, an initial release consisting of 4,654 math tutoring transcripts from a U.S.-based nonprofit online tutoring platform. MTM v1 serves as a first step toward a broader repository that is safe, open, large-scale, broad-coverage, and multimodal. By making tutoring interactions systematically observable and analyzable, MTM aims to support research on instructional processes, improve tutoring practice, and enable the development of AI systems grounded in real educational interactions.
\end{abstract}


\section{Introduction}
Tutoring is widely regarded as one of the most effective forms of educational support \cite{Dietrichson2021, Nickow2024, Robinson2021, VanLehn2011}. However, despite decades of research, our ability to systematically study tutoring at scale remains limited. A central bottleneck is data: while many platforms log student activity, far fewer capture the interactional dynamics of tutoring, including how tutors respond to students’ reasoning, confusion, and affect in real time.

Recent work has explicitly noted that the field still lacks large, reusable datasets of authentic tutoring interactions and that existing resources are often small, synthetic, or limited in pedagogical richness \cite{MacinaDaheimWang2023, Wang2023}. This gap constrains both scientific understanding of instruction and the development of AI systems that can meaningfully support teaching and learning.

To address this challenge, we introduce the Million Tutoring Moves (MTM) project. MTM aims to build a large-scale, open repository of tutoring interactions that captures tutoring as a dynamic, multimodal process. In this paper, we describe the motivation, design principles, and first release of MTM.

\section{Related Work}

Several datasets and research efforts have explored tutoring dialogue and instructional interaction. While these efforts have made important contributions, they also highlight key limitations that MTM is designed to address.

The CIMA dataset \cite{Stasaski2020} introduced a large open-access corpus of tutoring dialogues and explicitly noted the lack of large tutoring datasets. However, CIMA is constructed from scripted tutoring role-play from crowdsourced workers, which limits its ability to reflect the authenticity and variability of real-world tutoring.

More recent datasets such as TalkMoves \cite{Suresh2022} and MATHDIAL \cite{MacinaDaheimChowdhury2023} aim to incorporate richer pedagogical structure. These datasets improve grounding in real-world instruction, but remain relatively small in scale and often focus on only one subject.

A recurring challenge across these efforts is that authentic tutoring data is difficult to collect and share, due to privacy concerns, platform constraints, and the sensitive nature of educational interactions. The MTM dataset is designed to address this gap.

\section{MTM Design Principles}

We design MTM as a repository that is safe, open, large-scale, broadly representative, and ultimately multimodal, reflecting our goal of enabling rigorous and responsible study of tutoring interactions at scale.

\textbf{Safe}. We prioritize safety as a first-order design constraint. MTM is built on a rigorous de-identification pipeline that removes or transforms personally identifiable information while preserving the pedagogical and interactional structure of tutoring sessions. In particular, we adopt a “hidden in plain sight” strategy, in which sensitive entities are replaced with contextually appropriate surrogates rather than simply redacted. This approach maintains the natural flow and semantic coherence of dialogue—critical for studying instructional moves and discourse—while reducing the risk of re-identification. De-identification is complemented by systematic validation procedures to ensure both privacy protection and data utility.

\textbf{Open}. Subject to appropriate governance and access controls, MTM is designed as an open-access resource for researchers, educators, and developers. Our goal is to lower barriers to studying tutoring interactions by providing reusable datasets, standardized formats, and supporting infrastructure. By making MTM broadly accessible, we aim to enable reproducible research, facilitate cross-institutional collaboration, and support the development of educational technologies grounded in real-world data.

\textbf{Large-scale}. MTM is designed to support analysis at scale, both in terms of the number of tutoring sessions and the volume of interaction data within each session. While MTM v1, introduced below, already constitutes the largest publicly available tutoring interaction dataset to date (4,654 transcripts), it represents only an initial release, with the broader project aiming to scale to millions of tutoring moves across diverse contexts. This scale is essential for capturing variability in tutoring practices, enabling statistically robust analyses, and supporting data-hungry methods such as training modern large language models (LLMs).

\textbf{Broad-coverage}. Beyond scale, MTM emphasizes breadth of coverage across educational contexts. The repository is designed to span multiple subject areas, grade levels, and tutoring formats, including one-to-one tutoring, small-group sessions, chat-based interactions, video-based tutoring, and in-person settings. This diversity is critical for studying how tutoring strategies vary across domains and populations, and for developing models and insights that generalize beyond narrow settings.

\textbf{Multimodal}. Finally, MTM is designed with a long-term commitment to multimodal data. While MTM v1 focuses on text-based transcripts, the broader vision includes integrating additional modalities such as audio, video, and interaction traces (e.g., whiteboard activity). Multimodal data enables richer representations of tutoring, capturing not only what is said but how it is communicated, and how tutors and students interact with shared artifacts. This progression toward multimodality is essential for advancing both scientific understanding and AI systems that engage with learning in more human-like ways.

\section{MTM v1: 4,654 Math Tutoring Transcripts}

\subsection{Data Source and Educational Context}
MTM v1 consists of 4,654 math tutoring transcripts from a U.S.-based nonprofit organization that provides free, 24/7 online tutoring and college counseling to low-income middle and high school students. Students can connect one-on-one with certified academic coaches, often within 5–10 minutes of initial session request, and work in a virtual classroom that may include tools such as a whiteboard, document editor, text chat, and optional voice chat. The organization has a national footprint, with more than 25,000 students and over 100,000 on-demand tutoring requests served across all 50 states.

This context makes the dataset a strong starting point for MTM v1, which represents real-world tutoring at meaningful scale, serves an equity-oriented educational mission, and provides authentic interaction data from online tutoring sessions. For a first release, the focus on math transcripts offers a coherent instructional domain while still capturing a wide range of tutoring moves, problem-solving patterns, and student support strategies.

\subsection{De-Identification and Privacy Protection}
Given the sensitive nature of tutoring data, MTM v1 is built upon a rigorous de-identification pipeline designed to protect user privacy while preserving the pedagogical and interactional structure of tutoring sessions.

We develop an in-house pipeline for detecting personally identifiable information (PII) in tutoring transcripts with reference to major privacy protection guidelines, including the Children’s Online Privacy Protection Act (COPPA) \cite{FTC2013}, the Family Educational Rights and Privacy Act (FERPA) \cite{USDOE2011}, and the 18 enumerated personal identifiers that are required to be removed as valid de-identification according to the Safe Harbor approach provided by Health Insurance Portability and Accountability Act (HIPAA) \cite{USHHS2012}. The PII types considered in our current pipeline are \texttt{NAME} (excluding public figures or fictional names), \texttt{AGE}, \texttt{ADDRESS} (including state, town, street address, and zip code), \texttt{OTHER\_LOCATION} (local landmark names), \texttt{DATE}, \texttt{SCHOOL} (school names and numbering), \texttt{PHONE\_NUMBER}, \texttt{FAX\_NUMBER}, \texttt{EMAIL}, \texttt{SSN}, \texttt{ACCOUNT\_NUMBER}, \texttt{DEVICE\_IDENTIFIER}, \texttt{URL}, \texttt{IP\_ADDRESS}, and \texttt{OTHER\_IDENTIFYING\_NUMBER}.

Our PII detection approach combines two modern transformer-based sequence labeling models: \emph{DeBERTa-v3-base} \cite{He2020} and \emph{ModernBERT-base} \cite{Warner2025}, each fine-tuned on 1,181 tutoring transcripts annotated with PII ground truth (1,105 transcripts with 2,605,233 tokens for training and 76 transcripts with 408,124 tokens for validation) and both based on Transformers version 4.57.6. The final system aggregates predictions from both models to improve robustness and coverage, leveraging their complementary strengths in model architecture and parameter scale. Our PII detection approach achieves a precision of 0.9121, recall of 0.9773, and an F1 score of 0.9528 on an unseen test set comprising 100 tutoring transcripts (513,275 tokens).

Following detection, identified PII spans are transformed using a hidden-in-plain-sight (HIPS) strategy \cite{Carrell2013}. Instead of removing or masking sensitive information, we replace it with contextually appropriate surrogates that preserve the grammatical structure and semantic coherence of the dialogue.

This de-identification pipeline enables MTM v1 to balance strong privacy protection with high data fidelity, making the dataset suitable for open-access research while preserving the interactional richness required for studying tutoring processes.

\subsection{Scope of The Release}
MTM v1 is an initial textual release within a larger multimodal agenda. The broader MTM vision includes audio, video, and other traces, but this first version centers on transcripts in order to provide an immediately useful and lower-friction entry point for researchers. Even in transcript form, these sessions can support work on tutoring dialogue, instructional move detection, help-seeking, explanation quality, error remediation, and the development of tutoring taxonomies or AI annotation systems. MTM v1 is released alongside Sandpiper, an open sourced AI annotation application that allows researchers to annotate and study tutoring sessions \cite{Hedley2026}. This staged-release approach is also consistent with the NTO’s emphasis on safe processing, de-identification, and modular infrastructure.

\subsection{Descriptive Statistics}
We summarize the MTM v1 dataset using key descriptive statistics, including session counts, subject coverage, interaction structure, and lexical volume. The dataset contains 4,654 tutoring sessions spanning a broad range of secondary-level mathematics topics.

MTM v1 covers 15 mathematics subjects, ranging from middle school to advanced high school curricula:
\begin{itemize}
\item 6th-Grade Math, 7th-Grade Math, 8th-Grade Math,
\item Pre-Algebra, Algebra One, Algebra Two,
\item Geometry,
\item Trigonometry,
\item Statistics,
\item Pre-Calculus, Calculus AB,
\item Integrated Math One, Integrated Math Two, Integrated Math Three, and Integrated Math Four.
\end{itemize}

This diversity reflects the dataset’s broad coverage across grade levels and curricular structures.

\begin{figure}[htbp]
  \centering
  \includegraphics[width=.9\linewidth]{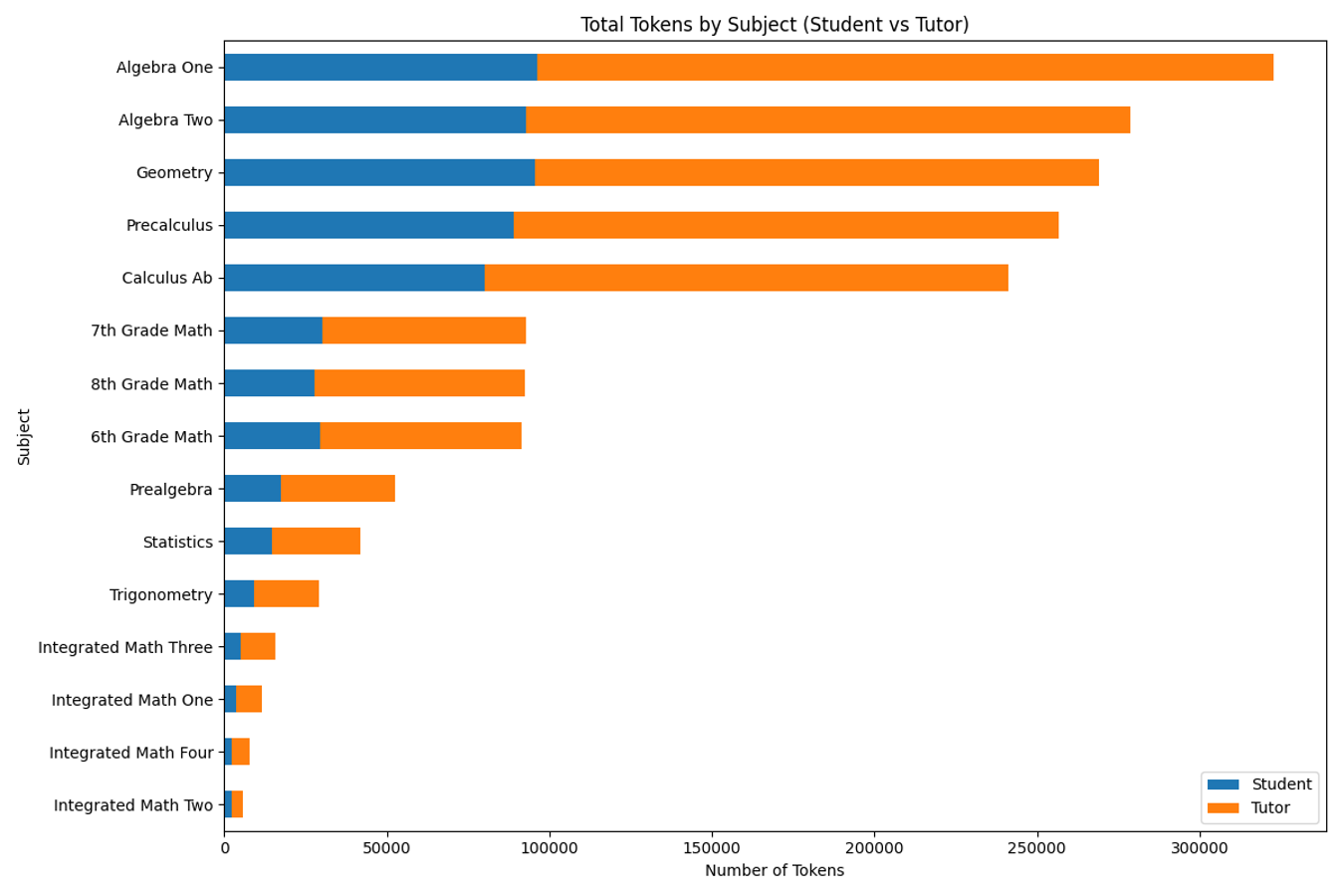}
  \caption{MTM v1’s Token Distribution by Mathematical Subjects and Roles}
  \label{fig:token_dist}
\end{figure}

MTM v1 captures over 4,000 hours of tutoring interactions, containing 274,121 utterances and 1,809,815 tokens in total. Table~\ref{tab:descriptive_stats} summarizes utterance-level and token-level statistics. Figure~\ref{fig:token_dist} shows the distribution of tokens across mathematical subjects and participant roles (student vs. tutor).

\begin{table}[h!]
\centering
\caption{MTM v1’s Utterance-Level and Token-Level Statistics}
\label{tab:descriptive_stats}
\begin{tabular}{lccc}
\hline
\textbf{Metrics} & \textbf{Total} & \textbf{Student} & \textbf{Tutor} \\
\hline
\multicolumn{4}{l}{\textbf{Utterances}} \\
\quad Total & 274{,}121 & 116{,}883 & 157{,}238 \\
\quad Average per session & 58.90 & 25.11 & 33.79 \\
\quad Median per session & 40 & 17 & 22 \\
\hline
\multicolumn{4}{l}{\textbf{Tokens}} \\
\quad Total & 1{,}809{,}815 & 595{,}617 & 1{,}214{,}198 \\
\quad Average per session & 388.87 & 127.98 & 260.89 \\
\quad Median per session & 264 & 85 & 168 \\
\hline
\end{tabular}
\end{table}

\section{Conclusion}
The Million Tutoring Moves project addresses a longstanding bottleneck in education research: the absence of open, large-scale datasets that capture tutoring as a rich, adaptive, human interaction. Situated within the NTO, MTM advances a vision of tutoring data that is simultaneously scientifically useful, practically relevant, multimodal, open, and safe. MTM v1, consisting of 4,654 math tutoring transcripts from UPchieve, is a first step toward that vision. By making tutoring interactions more observable and analyzable, MTM aims to support better research on teaching, better tools for tutoring organizations, and better AI systems for learning.

\bibliographystyle{plain}
\bibliography{references}

\end{document}